# FPGA-based electronic system for the control and readout of superconducting quantum processors

Yuchen Yang,[1,2] Zhongtao Shen,[1,2,a)] Xing Zhu,[3,a)] Ziqi Wang,[1,2] Gengyan Zhang,[3] Jingwei Zhou,[3] Xun Jiang,[3] Chunqing Deng,[3] and Shubin Liu.[1,2]

[1]*State Key Laboratory of Particle Detection and Electronics, University of Science and Technology of China, Hefei 230026, China*

[2]*Department of Modern Physics, University of Science and Technology of China, Hefei 230026, China*

[3]*Alibaba Quantum Laboratory, Alibaba Group, Hangzhou 310023, China*

[a)] Author to whom correspondence should be addressed: henzt@ustc.edu.cn and taoli.zx@alibaba-inc.com

Electronic systems for qubit control and measurement serve as a bridge between quantum programming language and quantum information processors. With the rapid development of superconducting quantum circuit (SQC) technology, synchronization in a large-scale system, low-latency execution, and low noise are required for electronic systems. Here, we present a field-programmable gate array (FPGA)-based electronic system with a distributed synchronous clock and trigger architecture. The system supports synchronous control of qubits with jitters of approximately 5 ps. We implement a real-time digital signal processing system in the FPGA, enabling precise timing control, arbitrary waveform generation, IQ demodulation for qubit state discrimination, and the generation of real-time qubit-state-dependent trigger signals for feedback/feedforward control. The hardware and firmware low-latency design reduces the feedback/feedforward latency of the electronic system to 125 ns, significantly less than the decoherence times of the qubit. Finally, we demonstrate the functionalities and low-noise performance of this system using a fluxonium quantum processor.

## I. INTRODUCTION

Quantum computing offers the possibility of a computational advantage for certain types of hard problems.[1, 2] Since the first demonstrations of the precise manipulation of a single quantum system,[3] a variety of quantum technologies have been investigated for the implementation of quantum computing.[4, 5, 6, 7, 8] Superconducting quantum circuits (SQCs) are one of the most promising quantum computing platforms.[9] The electronic system, for qubit control and measurement, serves as a bridge between the quantum programming language and the superconducting quantum processors. At present, the rapid development of SQC technology has entered a new stage — to scale up toward the demonstration of fault-tolerant quantum computing with logical qubits,[9] which puts forward new requirements for electronic systems.

Thus far, electronic systems with customized features are built and integrated with superconducting quantum processors with dozens of qubits.[10, 11, 12] Standalone control systems are also available as commercial products[13,14,15] are widely adopted by superconducting quantum computing research labs in academia. However, the technical details about the implementation of the aforementioned systems are often unavailable. Since a highly scalable and efficient quantum computing control system has to evolve as the progress of quantum processors, the development of such a system itself is a topic of active research regarding the architecture of quantum computers. Recently, open-source qubit controllers, which are based on the development board have been developed.[16,17] These systems are very flexible thus suitable for prototyping new quantum devices. To accommodate the need of both being scalable and applicable to fast real-time feedback/feedforward, field-programmable gate array (FPGA) based distributed control systems[18,19,20] are proposed and realized for superconducting qubit systems.

Here, we present a customized electronic system for the control and readout of superconducting quantum processors toward fault-tolerant quantum computing. Our system features a distributed architecture that is suitable for scaling up to a large number of qubits and a trigger and clock system specially designed for a low channel to channel jitter of approximately 1 ps and a feedback/feedforward latency as low as 125 ns. We structure this paper as followings: in Sec. II, we analyze the requirements of SQC system. In Sec. III, we discuss the implementation of an electronic control system in three parts: the clock and trigger system design, the low-latency design and the low-noise design. In Sec. IV, we experimentally characterize the performance of the electronic system. Finally, in Sec. V, we present qubit characterization data of a fluxonium superconducting processor obtained using the electronic system. In the appendices, we provide more details about requirements analysis and hardware implementation.

## II. REQUIREMENTS

In order to support the development of SQC towards fault-tolerant quantum computing[21], the electronic system should support high fidelity operations at a large number of qubits for quantum error correction (QEC).

### A. System

The basic idea of QEC is to define a logic qubit using a group of physical qubits according to some encoding scheme called quantum error correction code (QECC), and continuously correct errors that occur during computing.[22] Realizing QECC requires the electronic system to perform closed-loop feedback[23] to detect and correct the physical qubit errors in real time.

#### 1. Scalability

Surface codes is one of the most promising QECC. It has an error threshold of ~1% meaning that it can tolerate a physical error rate up to 0.01.[22] The number of physical qubits needed to define a logical qubit is strongly dependent on the error rates of the physical qubits. Although small distance surface code[24] is within reach of current superconducting circuit and control electronic technologies, there is no demonstration of reduced logical errors compared with the physical errors of a single-qubit. Therefore, we are motivated to design and build an electronic system for controlling the level of 100 qubits with the potential of testing QECC.

#### 2. Latency

Besides the number of qubits and the control fidelity which will be discussed in the following section, the efficiency of the QECC also depends on the feedback latency, which itself is a topic under active research. Reducing the ratio $r \equiv \tau_{FB}/T_1$ of the feedback latency $\tau_{FB}$ to the qubit lifetime $T_1$, can reduce the probability of state decay between the state detection and the feedback action. Since the probability of state decay is expected to be proportional to 1-exp(-$r$), for $T_1$ at commonly 100 μs level, a feedback latency of less than 100 ns would be needed under their feedback scheme to reduce the error probability to less than one part per thousand. Although different schemes of QECC have different feedback latency requirements, lower latency always improves the efficiency of QECC implementation[23] and reduces the probability of qubit state decay[25]. To further investigate the exact requirement of latency in QECC, it is desirable to able to reach a latency at the 100 ns level for performing feedback/feedforward algorithms in the context of dynamic quantum computing[12,13].

### B. High fidelity

Error rates substantially smaller than the QECC threshold allow smaller numbers of physical qubits.[22] Qubit errors are caused by a combination of many factors such as decoherence and imperfect control. Since the different types of imperfections cause different degrees of errors in different quantum algorithms or quantum gates, it is difficult to give these requirements for a general electronic system. We estimate the corresponding specifications at a given fidelity requirement when implementing specific quantum gates. Currently single-qubit gates are now routinely implemented with fidelity approaching 0.9999,[26] and two-qubit gates such as iSWAP can approach a fidelity of 0.999.[27] To avoid the electronic system becoming a bottleneck for gate fidelity improvement, we design the electronic system capable of reaching gate fidelities one order of magnitude higher than the current state of the art.

#### 1. Phase jitter

Jitter is one type of imperfections will introduce errors when applying control pulses on qubits. The system synchronization and clock performance mainly affect phase jitter of the control pulse. Phase jitter refers to the additional phase variation of the control pulse signal and it will deteriorate the phase accuracy of the control pulse. In this paper, we estimate the quantum gate error versus to the phase error due to phase jitter. Taking a pi-pulse as an example, for reaching 0.99999 fidelity, the requirement of the phase jitter is less than 6.2 ps. More details on this estimation are provided in Appendix A.

### 2. Spurious

For the control channel using sideband modulation, spurious signals come from the harmonic generation of the AWG and mixer leakage. The effect of spurious signals on the fidelity of quantum gates depends on many factors such as IF frequency $\omega_{IF}$ and the Rabi frequency $\Omega$ which is set by the signal amplitude. We model the spurious driving of a control signal and calculate the gate fidelity under the influence of spurious signals. Assuming that $\omega_{IF} = 100 MHz$, $\Omega = 33 MHz$, for reaching 0.99999 fidelity, the SFDR should be below -40dBc. More details on this calculation are provided in Appendix A.

### 3. DC bias noise

A high precision voltage generator is required to provide a DC bias for the qubit idling at its operation point. The DC bias noise at low-frequency acts as a quasistatic noise for qubit operation and contributes to the long-term drift of the devices. The high-frequency noise often has a white noise spectrum that would directly affect the decoherence of the qubit. High frequency noise is not a concern for DC bias as it can be removed by low pass filters with very a low cut-off frequency. Here, we calculate the noise specification for bias voltage generator (BVG) from the gate fidelity. The estimation depends on many factors such as the type of gates, qubit design and external circuits. We consider a case for iSWAP-like gate, for reaching 0.9999 fidelity, the BVG output voltage drift over a long period need to be less than 10 μV with a 5V output range. More details on this calculation are provided in Appendix A

## III. SYSTEM DESIGN

To deal with the increasing complexity of electronic system as the number of qubits increases, we have adopted a modular design approach. We modularize system different functions into TCM, AWG, DAQ, BVG, and reduce the coupling between each module as much as possible. If we built larger systems in the future, we just need to focus more on the inter-module relationship rather than its function details. The electronic system is integrated in 3U PXIe chassis and system communication is realized based on the PXIe protocol. Each module is interconnected with the high-speed backplane through high-density connectors, which greatly reduces system interconnection cables. Each module is implemented based on a FPGA, allowing specific functions to be realized while the backplane interface is compatible. Hence, different modules can be combined to adapt to different forms of SQC. Small-scale expansion can be achieved by inserting more modules inside the chassis. When the system scale reaches the level of about a dozen qubits, the system needs to be expanded to multiple chassis. As shown in Fig. 1, the PCIe communication from multiple chassis to the host is implemented by the MXI-Express system [28], and the clock synchronization of multiple chassis is ensured through the fan-out of the system root clock. We achieve synchronization of control and readout between multiple chassis through synchronization of trigger signals. On the basis of multi-chassis clock domain synchronization, we realize the synchronization of trigger signals by constructing a trigger daisy-chain architecture based on multiple TCMs as shown in Fig.2. In addition, Cross-chassis information transmission with low latency requirements is achieved by cabling high-density connectors on the front panels of each module.

The prototype architecture of a superconducting quantum computing system as an example is shown in Fig. 1. The electronic system includes a timing control module (TCM), four-channel arbitrary waveform generators (AWGs), four-channel data acquisition modules (DAQs), and six-channel bias voltage generators (BVGs). The AWGs have a 2-GSa/s sampling rate and a 14-bit amplitude resolution. The DAQs provide a 1-GSa/s sampling rate and 12-bit amplitude resolution. The BVGs provide an ultra-precise DC voltage. The TCM sends the system clock and global triggers to each module through a high-speed backplane to achieve the system level synchronization. For qubit control, the AWGs generate a control pulse sequence, which is then upconverted to the qubit frequency through a mixer. The pulses reach the qubits through microwave coaxial cables passing in a cryostat. Driven by microwave pulses, qubits undergo operations that enable universal operations for quantum computing. For qubit measurement, qubit states are obtained by sampling and decoding the measurement pulses that interact with superconducting resonators coupled to the qubits. The measurement pulses are generated by the AWG and upconverted by a mixer.

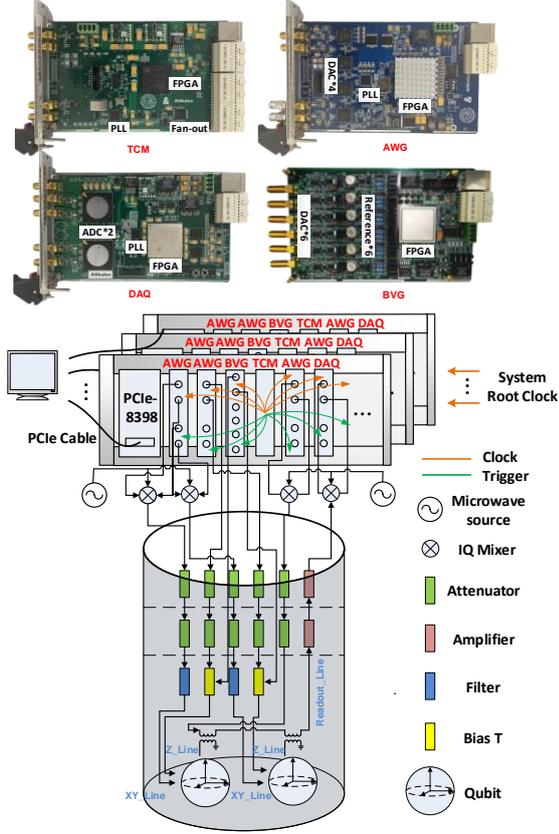

FIG. 1. System architecture. A prototype two-qubit system can be supported by two AWGs, a BVG, a TCM, and a DAQ.

## A. Clock and trigger system

With the increasing number of qubits, the electronic system presents a large-scale distributed characteristic. In such a system, the synchronization of each module greatly affects the stability and precision of qubit operations. In addition, the clock performance of AWGs and DAQs is positively correlated with the accuracy of qubit control and measurement.[29] To achieve system clock synchronization and improve clock performance, we propose a scalable high-performance clock system.

Each module requires clocks of different frequencies, including various analog clocks such as high-speed sampling clocks up to 2GHz, and digital clocks required by each module on FPGA. In order to ensure the synchronization of the system, all clocks must be related. We design a clock distribution network, as shown in Fig. 2. FS725 is used as the system root clock, which integrates a rubidium oscillator with ultralow phase noise. The 10 MHz root clock is sent to the phase-locked loop (PLL) chip LMK04821[30] on the TCM as the reference clock to generate two-channel 25 MHz clocks. One channel clock is sent to the FPGA on the TCM as the reference clock of the digital clock domain, and the other channel clock is fanned out to each module through the SY89829[31]. The fan-out clocks are routed on the high-speed backplane, and the skew between the clocks is kept as low as possible through equal-length differential pairs. In each module, the 25 MHz clock is used as the reference clock to generate the required clocks by the PLL chip or the digital PLL on FPGA. For example, in the AWG, four 2 GHz sampling clocks and one 25 MHz FPGA clock are generated by the PLL chip LMK04803[32]. Compared with the centralized clock scheme in which all clocks of the system are generated by a dedicated clock module, the distributed clock scheme simplifies the clock interconnection between modules and realizes the consistency of the clock interface of each module. This is consistent with the modular design idea described above, which is beneficial to improve the flexibility and scalability of the system. This distributed clock scheme needs to pay attention to two points to ensure the synchronization of the sampling clocks of each module:

1) The reference clock for each module cannot be divided before entering the phase detector of the PLL. The R1 coefficient in the PLL block diagram in Fig. 2 must be 1. The multi-phase problem from the clock divider will cause the clock phase relationship generated by the PLL of each module to be uncertain. Considering the frequency output capability of the TCM and the phase detection frequency range of the PLL1 on the AWG and DAQ, we chose 25 MHz as the frequency of the distributed clock.

2) PLL chips with zero-delay mode should be selected. Zero-delay mode established a fixed deterministic phase relationship between the phase of the reference clock and the phase of a feedback clock. Without using zero-delay mode, there are numerous possible fixed phase relationships from clock input to clock output.

A TCM can support clock synchronization scheme of up to 17 modules. When the system has further expansion requirements, the clock scheme needs to be further expanded. Fig.1 and Fig.2 show a system composed of multiple chassis, the system root clock is fanned out to TCMs in different chassis to achieve synchronization of multiple chassis. The clock skew due to cables can be adjusted by the PLL on the TCM and the synchronization between the chassis is guaranteed by the PLL with zero-delay mode. Synchronization within a chassis is still guaranteed by a TCM-based clock and trigger scheme.

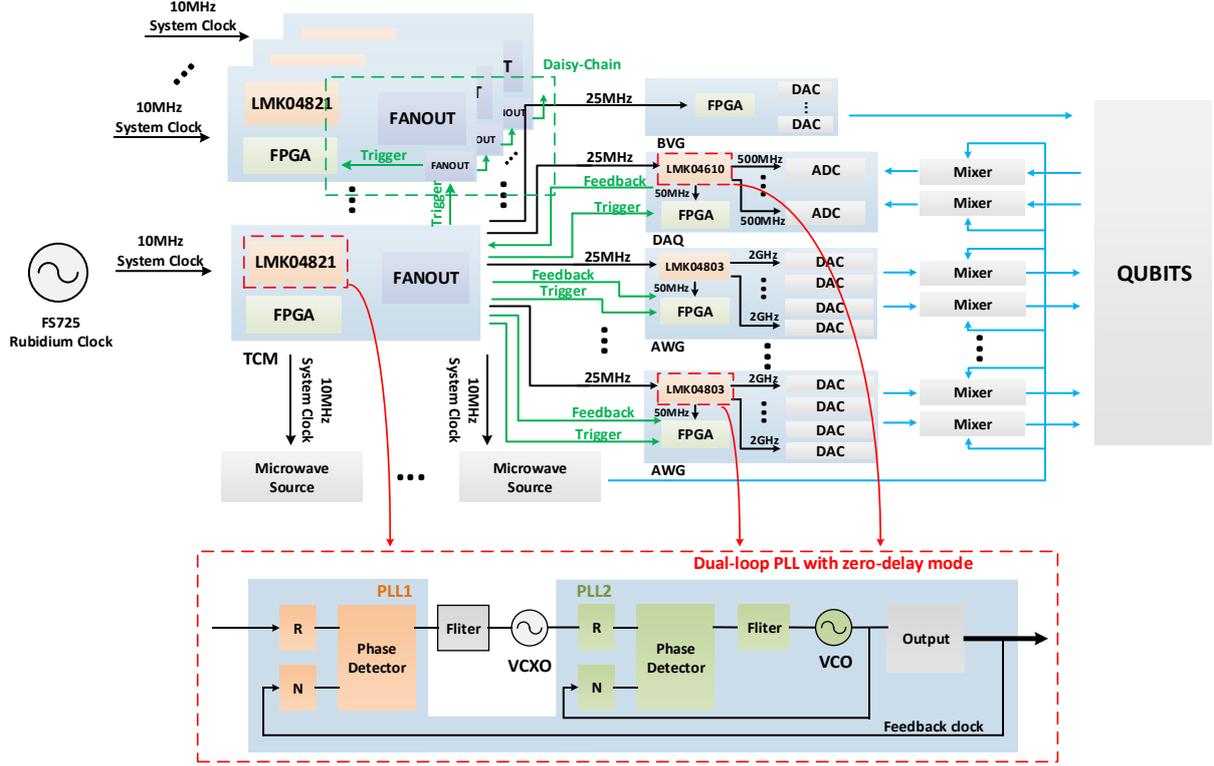

FIG. 2. Clock and trigger system. The system clock (10 MHz) is locked to 25 MHz in the TCMs by PLLs (LMK04821) and fanned out to each module as the synchronous reference clock. The 25 MHz reference clock is locked to 2-GHz sampling clocks via PLLs (LMK04803) in each AWG. In the DAQ module, the reference clock is locked to 500 MHz sampling clocks by PLLs (LMK04610) [33]. The 10 MHz root clock is used as the reference clock for the microwave source, which is applied to generate the local oscillator (LO) signal. The red box is a typical dual-loop PLLs with zero-delay architecture. The green box is a trigger daisy-chain architecture based on multiple TCMs.

To improve clock performance, we used FS725 as the system root clock, which integrates a rubidium oscillator with ultralow phase noise. In addition, the PLLs for clock synchronization on TCM, AWGs, and DAQs have a dual-loop architecture, which can provide low jitter performance over a range of output frequencies and phase noise integration bandwidths. PLL1 uses a narrow loop bandwidth to retain the frequency accuracy of the reference clock input signal while simultaneously suppressing the higher offset frequency phase noise that the reference clock may have accumulated along its path or from other circuits. The low phase noise reference provided to PLL2 allows PLL2 to operate with a wide loop bandwidth. The loop bandwidth for PLL2 is chosen to take advantage of the superior high offset frequency phase noise profile of the internal VCO and the good low offset frequency phase noise of the reference VCXO. Ultra-low jitter is achieved by allowing the external VCXO phase noise to dominate the final output phase noise at low offset frequencies and the internal VCO's phase noise to dominate the final output phase noise at high offset frequencies.

In quantum computing experiments, the electronic system completes a series of control and read operations with strict timing according to the upper-level algorithm. The clock system ensures that each module has a synchronous digital processing clock and synchronous sampling clock. On the basis of clock synchronization, we propose an FPGA-based trigger system. The trigger generated by the TCM is defined as level 1, which is a synchronization trigger that starts a complete operation of qubits (one lifetime cycle of qubits). It is transmitted to each AWG and DAQ through the backplane by a star connection architecture. We define the trigger generated by each module as level 2, which controls the timing of each module's waveform output and waveform sampling. Fig. 3 shows how the trigger system works with a simple qubit control and readout example.

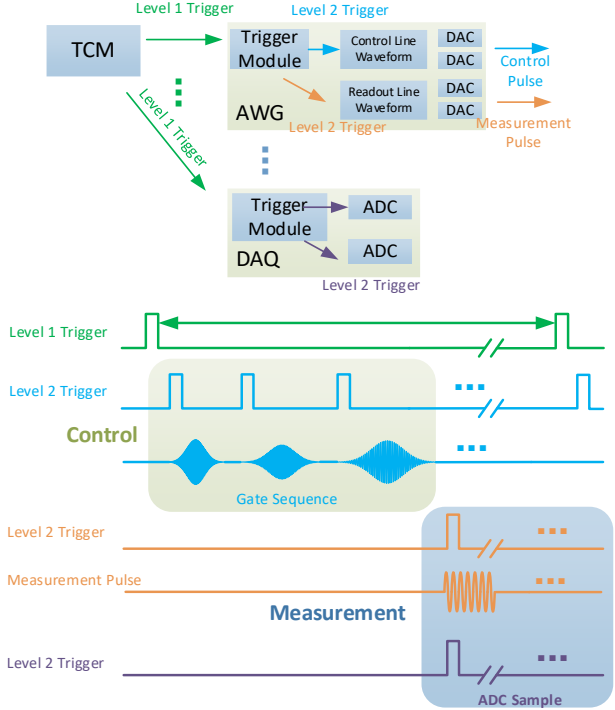

FIG. 3. Timing of the multilevel trigger architecture. The TCM decomposes the tasks into a series of real-time task sequences and sends trigger sequences to each module. The AWGs and DAQs generate different level 2 trigger sequences according to the system trigger and configuration information (implemented in the FPGA). The DACs and ADCs emit gated sequence waveforms and sample the probe signal carrying the qubit status information according to the level 2 trigger, thus realizing control and measurement of a qubit

## B. Low-latency design

Realizing QEC requires the electronic system to perform closed-loop feedback to detect and correct the physical errors in real time. The lower latency of the feedback means that fewer errors are likely to occur due to decoherence during the wait time and thus a higher QEC efficiency.

As shown in Fig. 4, we define the latency $\tau_{FB}$ of the feedback loop as the time from the beginning of the measurement pulse until the completion of the feedback control pulse.

$$\tau_{FB} = \tau_{EL} + \tau_{RO} + \tau_{CP}$$

where $\tau_{RO}$ is the readout duration, $\tau_{CP}$ is the duration of the control pulse, $\tau_{EL}$ is the total electronic system delay. The propagation delay through the microwave coax connecting to the cryogenic setup, the readout duration $\tau_{RO}$ and the duration of the control pulse are system-specific. The electronic feedback latency is more concerned with the response delay caused by the AWG and DAQ, and the system-specific delay are generally excluded. This chapter focuses on how to reduce this latency from hardware and firmware design.

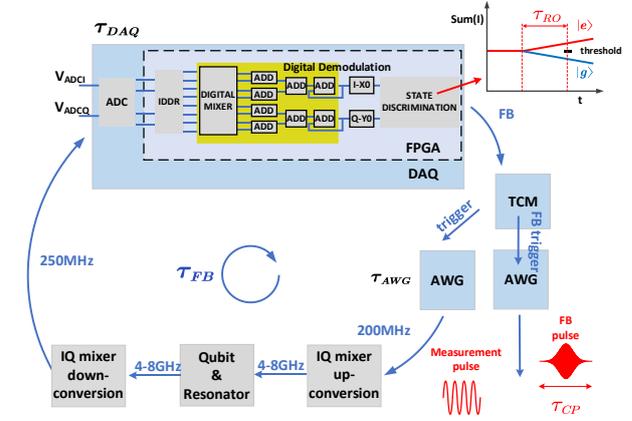

FIG. 4. Overview of the feedback loop. The feedback loop starts from the beginning of the measurement pulse to the generation of the next FB (feedback) pulse conditioned on the measurement. We define the readout duration $\tau_{RO}$ as the integration time of the measurement pules.

### 1. Hardware

In the hardware circuit, the data transmission time between the FPGA and ADC/DAC is a contributor to the latency. There are currently two mainstream digital interfaces for high-speed data transmission, JESD204B and LVDS. The JESD204B interface performs operations such as scrambling and encoding to achieve a higher transmission rate. Compared with the JESD interface, the LVDS interface is based on the data transmission of the physical layer with lower timing overhead. To achieve lower latency, we used 112 pairs of LVDS and 48 pairs of LVDS to implement data transmission between FPGA and DACs/ADCs on the AWG and DAQ. The utilization of the LVDS interface contributes to a latency time within a few nanoseconds.

The LVDS interface has the advantage of low latency, but due to the large number of high-speed parallel differential pairs, the timing requirements for data transmission are higher. The synchronization clock system and the device clock-loop architecture of DAC/ADC and FPGA ensure that the clocks at both clocks at both ends of the LVDS interface have a definite timing relationship. Simultaneously, the phase relationship between the data and clock can be adjusted through the DELAY module to ensure that the data meet the setup and hold time requirements.

On the readout line, we performed IQ analog downmixing on the microwave, which carries qubit state information.

### 2. FPGA Firmware

In AWG digital signal processing (DSP), the waveform generator module is the main latency contributor. The waveform data sent by the host are prestored in memory and are loaded to the DAC when the trigger comes. We have used block random-access memory (BRAM) instead of double-data-rate fourth generation (DDR4) to prestore waveform data. Compared with each other, BRAM integrated inside the FPGA has a lower readout latency (2 clock cycles vs. 14 clock cycles), but the storage space is smaller (30 Mb vs. 4 Gb). To generate long waveforms relevant to quantum computing, we need to use storage space more efficiently. Analyzing the waveform data, we found that it has low entropy characteristics, i.e. a large portion of a waveform is occupied by zero data representing idle time and repeated pulse waveforms data. This means that the waveform data have great compression potential. Based on the flexible trigger system design, we developed the waveform sequence generation function using the level 2 trigger, as shown in Fig. 3. A long waveform representing a gate sequence can be can be decomposed into pulses starting at given delay times, so that only a number of pulse envelopes corresponding to the gates in a finite group and the start time of each of their appearance need to be stored. This waveform generation module using only on-chip BRAM is capable of performing common qubit experiments such as the characterization of coherence times, gate calibration, randomized benchmarking and etc.

On FPGA integrated in the DAQ, we implemented a pipelined DSP circuit under the 250-MHz clock, as shown in Fig. 4. We realize digital demodulation through I and Q digital mixing followed by an accumulator module to improve the signal-to-noise ratio (SNR). In the digital mixer, the input signals $V_{ADCI}$ and $V_{ADCQ}$, as defined in Eq. (1), are multiplied by a complex exponential of $\omega_{IF}$ to realize digital downconversion. The complex output signal $S(t)$ is obtained as:

$$S(t) = (V_{ADCI} + V_{ADCQ} * j) * e^{-j w_{IF} n / f_s}$$
$$= \left(V_{ADCI} * \cos(w_{IF} n / f_s) + V_{ADCQ} * \sin(w_{IF} n / f_s)\right) +$$
$$j * \left(V_{ADCQ} * \cos(w_{IF} n / f_s) - V_{ADCI} * \sin(w_{IF} n / f_s)\right).$$

In the FPGA implementation, we use $f_s/4$ ($f_{IF}$ =250 MHz) as the IF frequency. Where $w_{IF} = 2\pi f_{IF}$, $\cos(w_{IF} n / f_s) = \cos(\pi n/2)$, the multiplication in Eq. (1) can be optimized. Therefore, the digital mixer structure has the lowest latency (1 clock cycle = 4 ns) because no multiplier is needed.[25] After the digital mixer, we use a three-stage adder to realize the accumulation of data. With the accumulation of data, the SNR of the demodulation result is improved, and the accuracy of the qubit state measurement is improved. The state discrimination module compares the rotated data to the threshold to determine the qubit state. Depending on the state, the DAQ sends a feedback signal to the TCM. In addition, the DAQ can upload the measurement results or raw data to the host through PCIe.

In general, we reduced the pipeline latency of each stage to 4 ns by increasing the FPGA operating frequency. At the same time, a lower latency waveform generator module and digital demodulation module are implemented in FPGA. The DSP latencies of AWG and DAQ are approximately 16 ns (4 clocks) and 20 ns (5 clocks), respectively.

### C. Low-noise design

Qubits are quite sensitive to noise, and the control and measurement of qubits with high decoherence time requires that the noise of the electronic system be sufficiently low. As shown in Fig. 1, the noise of the electronic control system is mainly coupled to the qubit through the readout line, XY line and Z line. The noise of the readout line and XY line mainly includes the spurs and harmonics generated by AWGs and leakage from mixers. The noise on the Z line is a sum of the noise from DAC, amplifier and reference voltage on the BVGs.

The noise on AWGs depends on the performance of the DAC chip and the quality of the sampling clock. Considering the sampling rate, bandwidth, LVDS interface and noise level, we chose the AD9739[34] chip, which is a 14-bit, 2.5 Gsps high-performance DAC. To improve the quality of the clock, the 25 MHz reference clock is locked to 2-GHz sampling clocks via LMK04803, which is an ultralow noise clock jitter cleaner with dual-loop PLLs. In addition, the output of DAC is connected with a low-pass filter to further reduce out-of-band noise.

The BVG provides a precise DC bias for the qubit work at any operation points. Compared with the AWG

working in the ultralow duty cycle pulse mode, the BVG always provided the dc bias required by the qubit during the experiment, and the qubit frequency and decoherence were highly sensitive to the dc bias. Therefore, the bias voltage noise of the BVG needs to be extremely low (less than 10 µV). To achieve such a low noise level, the design is based on a 20-bit DAC (AD5791)[35], ultralow noise and low-temperature drift amplifiers, reference voltage circuits, and low-temperature-coefficient resistors. We characterize the low-frequency noise to evaluate the DC performance of the circuit. In the 0.1–10 Hz bandwidth, the DAC generates approximately 0.6 µVp-p noise, the amplifiers (AD8675)[36] generate 0.66 µVp-p noise, add this to the 1.2 µVp-p noise from the voltage reference (LTZ1000ach)[37], and the expected output noise is approximately 1.6 µV p-p, which meets our requirements.

Due to the mixer's DC offset and the imbalance of the IQ analog channel, the output signals from the mixer have LO leakage and sideband leakage. We perform digital precompensation processing on the waveform to suppress the leakage from the mixer.

For the implementation detail of the AWG, DAQ, and BVG, please see Appendix B.

## IV. TESTING

In this section, we present the test results of the electronic system.

### A. Jitter

We usually use skew and jitter to characterize the synchronization of the system. The skew is determined by the error of the electrical connection length, which is a constant for the system and can be compensated. The impact of jitter cannot be calibrated, so we are more concerned with the jitter performance of the system. We measured the channel to channel jitter and phase jitter of the AWG output.

To measure the channel to channel jitter, the two AWGs' output channels were connected to the DAQ input, and repeatedly sample the two channels output 5000 times with a time interval of 100 us. Through fixed-point phase analysis of each sampled waveform, the jitter of the channel to channel can be calculated. Since the input bandwidth of DAQ is 4.5MHz~400MHz, this can be regarded as the integration bandwidth of position jitter measurement. Fig. 5 shows the jitter among two AWGs' output channels. The standard deviation of AWGs' output channel to channel is approximately 5 ps. Low jitter means that the system synchronization scheme is capable of supporting high-precision synchronized qubit operations.

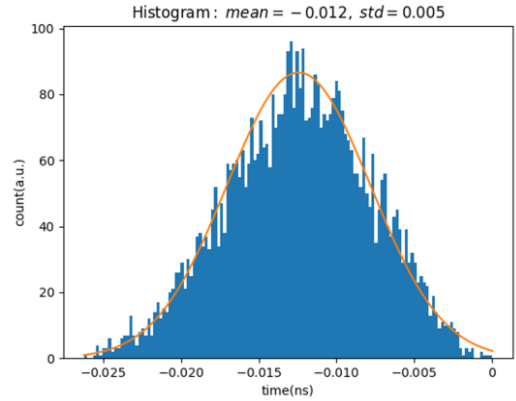

FIG. 5. The histogram of the channel to channel jitter of two AWGs.

To measure the phase jitter, we measured the phase noise of the AWG output at five frequencies, i.e., 50, 100, 200, 300 and 400 MHz as shown in Fig. 6. The 100MHz RMS phase jitter with 10Hz~30MHz integration bandwidth is about 1.8 ps, which is meet the requirement of 6.2 ps.

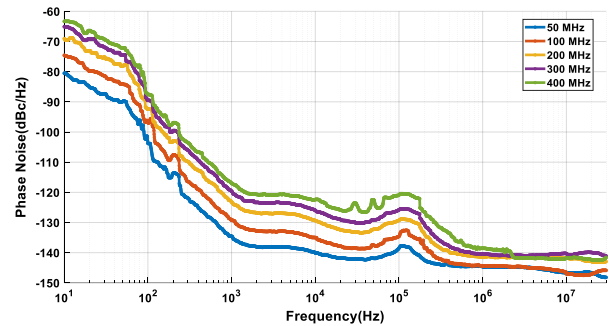

FIG. 6. Phase noises at different frequency of the AWG output.

### B. Noise

#### 1. Arbitrary waveform generator

The spurious-free dynamic range (SFDR) is a crucial specification that can be used to characterize the dynamic performance of AWGs. SFDR specifies the relationship between the amplitude of the fundamental frequency being generated and the amplitude of the most prominent harmonic. The AWG output spectrum was measured by a Keysight N9010A spectrum analyzer with a frequency range of DC~2 GHz. The results of a 400-MHz output signal are shown in Fig. 7. Table I reports the SFDR at six frequency points. We achieve an average SFDR smaller than -60 dBc over the output band, which is meet the requirement of SFDR should be below -40 dBc.

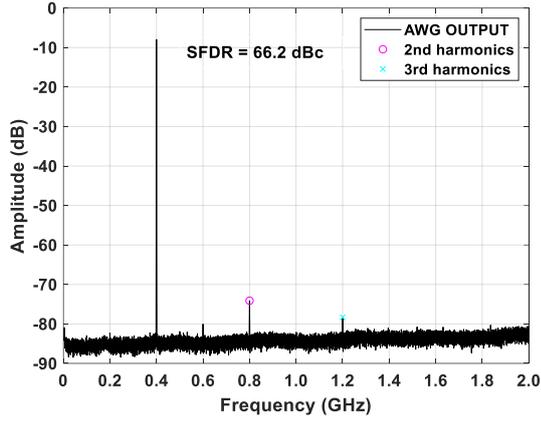

FIG. 7. Spectrum of the AWG output at 400 MHz.

TABLE I. SFDR test results.

| Frequency (MHz) | SFDR (dBc) |
|---|---|
| 10 | -70.8 |
| 100 | -60.2 |
| 200 | -55.7 |
| 300 | -58.1 |
| 400 | -66.2 |
| 500 | -64.5 |

### 2. Data acquisition module

The DAQ module was tested according to IEEE standard 1241-2010.[38] We used a radio frequency signal generator (SMA100) to generate the test signals. Figure 8 shows the spectrum of 398 MHz signals acquired by the DAQ. Table II presents the DAQ test results (where SNR denotes the signal-to-noise ratio, THD is the total harmonic distortion, and ENOB is the effective number of bits).

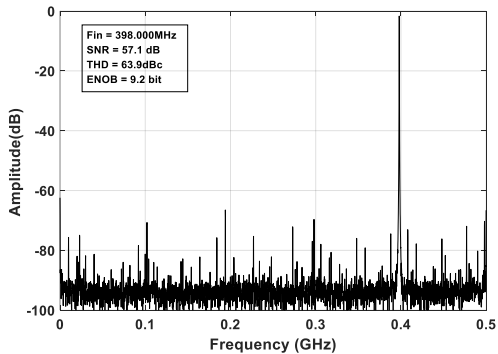

FIG. 8. Spectrum of a 398 MHz signal acquired by the DAQ.

TABLE II. DAQ test results.

| Frequency (MHz) | SNR (dB) | THD (dBc) | ENOB (bit) |
|---|---|---|---|
| 19.9 | 60 | -65.5 | 9.4 |
| 49 | 58.3 | -65.0 | 9.3 |
| 98 | 58.2 | -65.6 | 9.3 |
| 148 | 58 | -62.9 | 9.2 |
| 198 | 58.1 | -64 | 9.2 |
| 248 | 57.4 | -61.9 | 9.1 |
| 298 | 57.1 | -59.6 | 9.1 |
| 348 | 57 | -57.1 | 9.0 |
| 398 | 57.1 | -63.9 | 9.2 |

### 3. Bias voltage generator

The output noise of the BVG is sensitive to environmental temperature changes. We test BVG in an A/C laboratory. The output noise of the BVG was tested using a 6 $\frac{1}{2}$ Digital Multimeter (DMM, Fluke 2638A). The 10-h output values at 0.86 V are shown in Fig. 9, where Vpp is approximately 6 μV, which is meet the requirement of BVG's output voltage drift over long-term (10h) less than 10 μV.

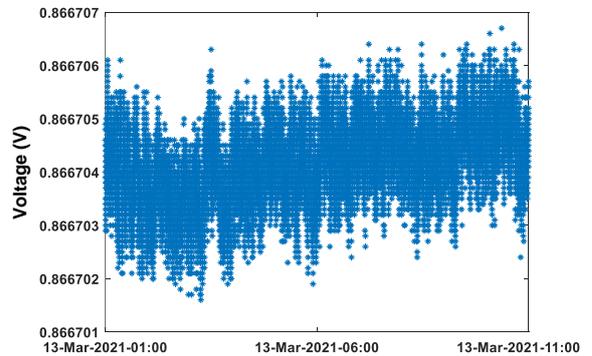

FIG. 9. +0.86 V output over 10 h from Fluke 2638A.

### 4. Leakage

The mixer output spectrum was measured by a Keysight N9010B spectrum analyzer. As shown in Fig. 10, the LO leakage and sideband leakage can be suppressed to approximately -50 dBc, which is meet the requirement of SFDR should be below -40 dBc.

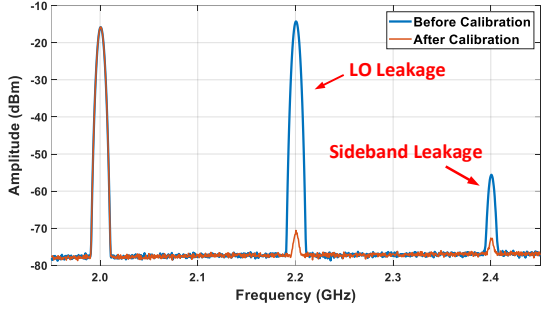

(a) Comparison of spectrum (2 GHz output) before and after calibration

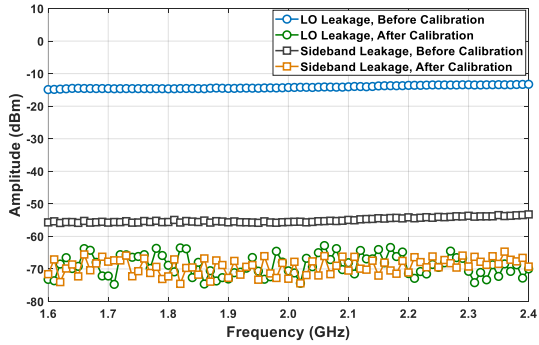

(b) Calibration performance at different frequency points.

FIG. 10. Mixer calibration performance.

### C. Latency of feedback

To test the implementation of real-time demodulation and measure the closed-loop feedback latency of the electronic system, we built a testing platform, as shown in Fig. 11.

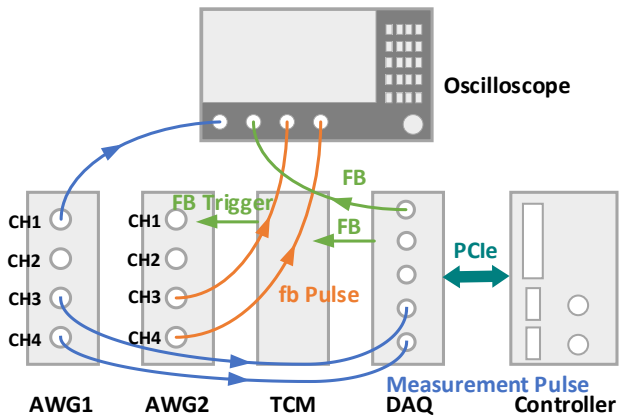

FIG. 11. Block diagram of the testing platform. CH3 and CH4 of AWG1 generated a measurement pulse containing qubit state information and sent them directly to the DAQ, while CH1 generated the same waveforms and was connected to an oscilloscope for observation. The DAQ sampled the pulse signal and generated the corresponding FB (feedback) signal after digital processing. The FB is sent to TCM and the oscilloscope. CH3 and CH4 of AWG2 generate an FB pulse based on the FB trigger from TCM and send it to the oscilloscope.

Based on the testing platform, we measure the DSP latency in DAQ and the electronic system latency. To observe the timing of signal processing stages in DAQ, the parallel data in the FPGA were uploaded to the controller via PCIe, and a parallel-to-serial conversion was performed in the software. The signal at different processing stages is shown in Fig. 12. Blue squares (ground state) and red squares (excited state) represent the corresponding digital signals obtained from the FPGA design. Digital processing takes a time of $\tau_{PROC}$ = 20 ns. The electronic system latency sampled by the oscilloscope is shown in Fig. 13. The measured electronic system feedback latency is approximately 125 ns (excluding the readout time of 48 ns).

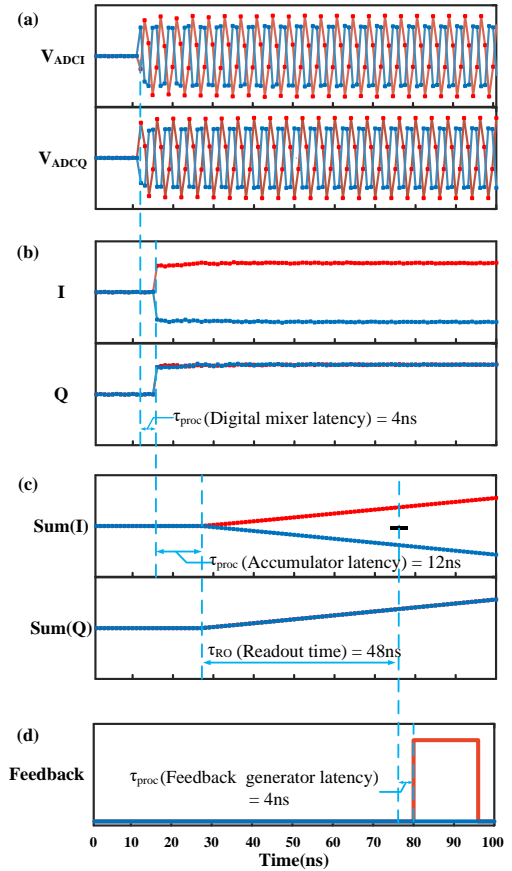

FIG. 12. Calculated signals at different stages for exemplary inputs (ground state: blue line; excited state: red line). (a) The signals $V_{ADCI}$ and $V_{ADCQ}$ from the ADC, which are the digitization of the IQ signals of the intermediate frequency. (b) In-phase (I) and quadrature (Q) components of the signal obtained at the output of the digital mixer. (c) Output of the accumulator. (d) The generation of the feedback signal depends on the output of the state discrimination module.

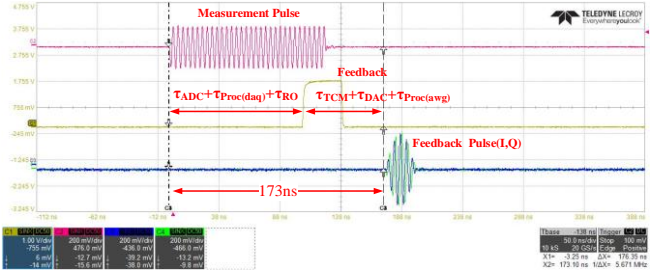

FIG. 13. Test results for the feedback latency.

## V. QUBIT EXPERIMENTS

To demonstrate the performance of this control and readout electronics system, we performed qubit characterization using a fluxonium superconducting processor.[39] A picture of the experimental setup is shown in Fig. 14 (a). Here, we present data from one-tone spectroscopy, two-tone spectroscopy, relaxation time $T_1$ measurement, and Ramsey dephasing time ($T_2^*$) measurement.[40]

We performed readout cavity spectroscopy by sweeping the AWG1 frequency and BVG voltage. The data are shown in Fig. 14(b) (c). We performed qubit spectroscopy by sweeping the AWG2 frequency and BVG voltage. The data are shown in Fig. 14(d) (e). We measure a qubit minimum frequency approximately 1.82 GHz when the qubit is biased at its half-flux quantum point. To characterize the lifetime of the qubit, we measured the relaxation time $T_1$ and the dephasing time $T_2^*$. The measurement was performed by preparing the qubit in the excited state with a π-pulse and waiting for a variable time t$_{wait}$ to measure the qubit state (Fig. 14(f)). Ramsey measurement was performed by preparing the qubit in the superposition state with a $\pi/2$-pulse and waiting for a variable time t$_{wait}$ to apply another $\pi/2$-pulse and measure the qubit state (Fig. 14(h)). Fig. 14(g) and Fig. 14(i) We obtain qubit $T_1$ = 90 μs and Ramsey $T_2^*$ = 19 μs.

## VI. CONCLUSIONS

We have described the design and implementation of a high-performance electronic system for the control and readout of superconducting quantum processors. The distributed system architecture is suitable for scaling up to a large number of qubits. The system supports synchronous control of qubits with jitters of approximately 5 ps. The system also enables real-time analysis of the qubit state. The low-latency design enables a feedback latency of the electronic system as low as 125 ns, significantly less than the decoherence times of the qubit. The achieved system-level synchronization and feedback latency reach advanced levels.[18, 19, 20, 25, 41] We have used this system to characterize a fluxonium superconducting processor. We obtained $T_1$ = 90 μs and $T_2^*$ = 19 μs. Together with our noise measurement of the electronics and analyses, we show that this control system is capable for high fidelity qubit operations. We proposed and realized a high-performance customized control electronic system that can support fault-tolerant quantum research and provide a reference solution for researchers trying to develop an electronics platform in their lab.

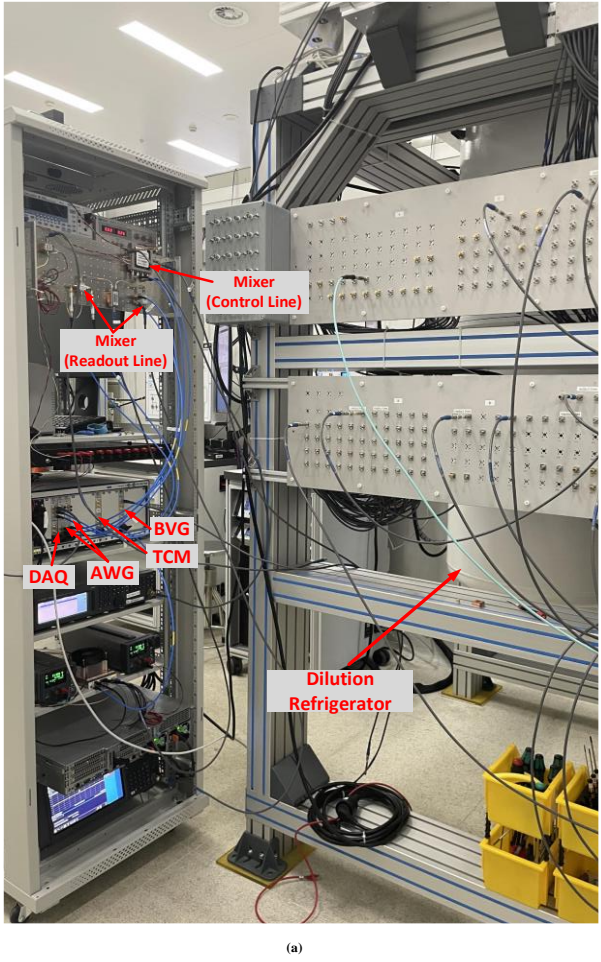
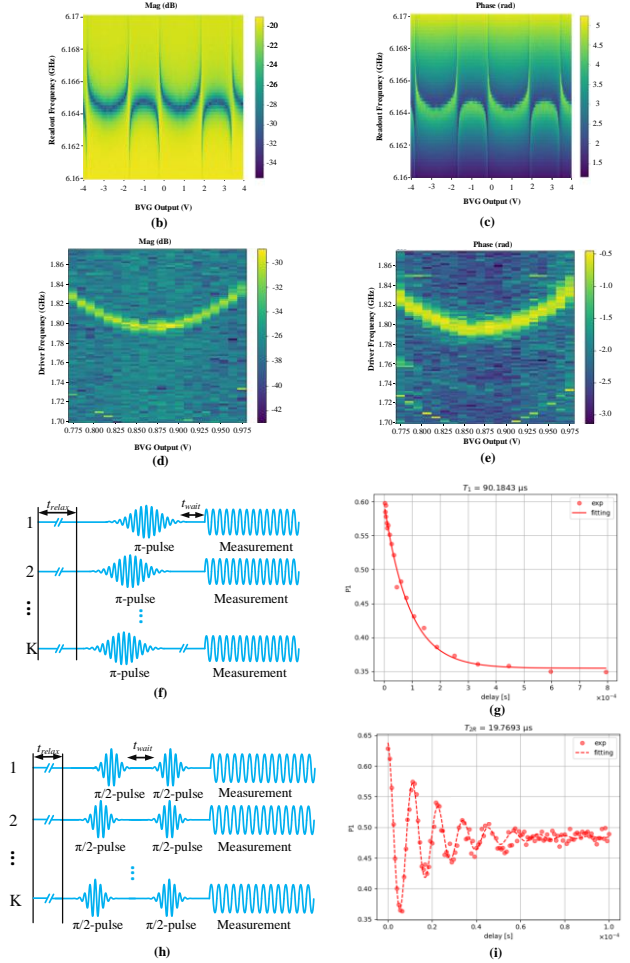

FIG. 14. The electronics system was used to characterize a fluxonium superconducting qubit. (a) Picture of the experimental setup. (b) Amplitude-frequency response of the readout cavity with different BVG output voltages. (c) Phase-frequency response of the readout cavity with different BVG output voltages. (d) Amplitude-frequency response of a qubit with different BVG output voltages. (e) Phase-frequency response of a qubit with different BVG output voltages. (f) $T_1$ measurement timing. (g) Qubit $T_1$ measurement data. (h) $T_2^*$ measurement timing. (i) Qubit $T_2^*$ measurement data.

## Appendix A: Requirements Analysis

### 1. Jitter

For a single-qubit rotation defined as

$$U(\theta, \varphi) = \exp[-i\frac{\theta}{2}(\cos\varphi\,\sigma_x + \sin\varphi\,\sigma_y)],$$

the rotation angle θ is controlled by the amplitude and duration of the resonant driving signal and the angle $\varphi$ between the rotation axis and x-axis in the Bloch sphere is controlled by the phase of the signal. The jitter from phase noise can be considered an error in the angle of the rotation axis $\varphi$.

We calculate the gate error of such rotations with respect to a standard π-rotation around the x-axis, $U(\pi, 0)$. The gate fidelity of a given unitary $U_1$ with respect to a target unitary $U_2$ is defined as

$$F_g = \frac{\text{Tr}[U_1 U_2^\dagger U_2 U_1^\dagger] + \left|\text{Tr}[U_2^\dagger U_1]\right|^2}{d(d+1)}$$

with $d = 2$. To reach a gate fidelity of 0.99999 for a π-rotation, we require the error in $\varphi$ smaller than 0.00387 rad. This phase error translates into a jitter of 6.2 ps assuming a 100-MHz IF signal is generated.

### 2. SFDR

We model the drive of a spurious signal that has an amplitude of $m$ times the amplitude of the main control signal. The main control signal is for a single-qubit gate define as

$$U = \exp(-i\theta\sigma_x/2),$$
a rotation around x-axis by θ. If we assume a Rabi frequency of Ω, IF frequency $\omega_{IF} = 2\pi \times 100$MHz, the gate time is $t = \theta/\Omega$. With the spurious drive, the qubit undergoes a rotation given by
$$U_1 = \exp(-i(\omega_{IF}\sigma_z/2 + m\Omega\sigma_x/2)t).$$
For an identity gate without driving, we have
$$U_2 = \exp(-i(\omega_{IF}\sigma_z/2)t).$$

We calculate the gate fidelity vs SFDR assuming the maximum signal corresponds to $\theta = \pi$ for a gate time normally at 10 ns, $\Omega = 2\pi \times 50$ MHz. For reaching 0.99999 fidelity, SFDR should less than -40 dBc.

### 3. DC bias noise

For the gate fidelity estimation, we consider a "typical" case of an iSWAP-like gate[39]. This consideration is not general but we can probably treat it as a worst-case scenario as such gate based on resonant interaction is usually the most susceptible to control errors. For reaching 0.9999 fidelity, we need a precision of 1e-5 flux quantum on the flux bias. Then we can convert the required 1e-5 flux quantum precision to the electronics precision by considering the following bias line configuration. To isolate the qubit from noise at room temperature, we typically have R=1k ohm and M = 2pH to make sure we can tune the device through multiple flux quantum within a 5V range. The quasistatic voltage noise required is
$$\delta V = 10^{-5}\Phi_0 R/M \approx 10\mu V$$
We require the voltage drift over long term (10h) within Vpp <10 μV so it can maintain a high fidelity over long time without recalibration

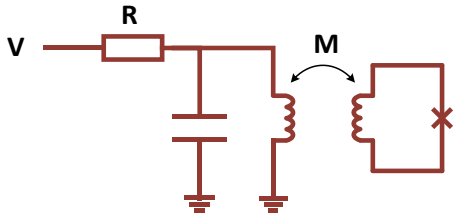

Fig. 15. Typically flux bias circuit.

## Appendix B: Implementation details

### 1. Arbitrary waveform generator

The block diagram of the AWG is shown in Fig. 16. A Xilinx Ultra-scale Kintex060 FPGA is utilized as a processing central. The AWG is implemented with four DACs (AD9739), which provide a 2-GSa/s sampling rate and a 14-bit resolution. Communications based on PCIe protocol between the host and the hardware are handled by a PCIe IP CORE on FPGA. The dual loop PLLs LMK04803 is a high-performance clock conditioner utilized to manage clock jitter cleaning, synchronization, generation, and distribution. A "direct mode" waveform generation module is implemented on FPGA for lower latency. The waveform data sent by the host is prestored in BRAM and can be directly sent to DAC without calculation. We use 112 pairs of LVDS to realize the data transmission between FPGA and four DACs. On the FPGA side, we use 112 parallel OSERDES3 to achieve high-speed data serialization (4:1). OSERDESE3 in UltraScale devices is a dedicated parallel-to-serial converter with specific clocking and logic features designed to facilitate the implementation of high-speed source-synchronous applications. Considering the Z line control and the mixer's DC offset correction, the DAC output is connected to the high-speed amplifier LMH6703 to achieve DC coupling. In addition, the output of DAC is connected with a low-pass filter (LFCN-530) to further reduce out-of-band noise. It should be pointed out that due to the state of the divide-by-four circuit inside the DAC is unknown at power-up, the resulting phase ambiguity will cause a ±2 sampling clock offset between multiple DACs. We realized the synchronization of multiple DACs on one board through the solution provided in Ref. 18. DAC synchronization between multiple boards is achieved through a power-on calibration that involves external measurement instruments.

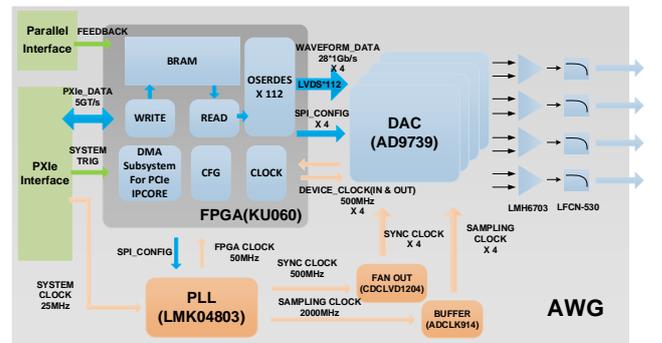

Fig. 16. Block diagram of AWG.

## 2. Data acquisition module

The block diagram of DAQ is shown in Fig. 17. The DAQ is implemented with two ADCs (ADC12D1000) [42], which integrated two ADCs with 1-GSa/s sampling rate and 12-bit resolution. Considering that 10 qubits are generally coupled on a readout line, and the bandwidth of a single qubit is typically 30MHz, the instantaneous bandwidth of the readout line is about 300MHz. The 1-GSa/s sampling rate can meet the requirements of the instantaneous bandwidth. The analog IQ signal is RF coupled to the ADC through the balun, and then digitized and transmitted to the FPGA through 96 pairs of LVDS. In digital signal processing, we simplify the digital demodulation process by fs/4 sampling, which reduces the latency of digital mixing, but this is only valid for signals with an intermediate frequency of fs/4. Additional processing is required in multi-channel demodulation, the structure of parallel multi-channel demodulation is shown in Fig.18. Double decimation of the fs/4 demodulated IQ signal, which can reduce the number of multi-channel parallel DDS and multipliers by half.

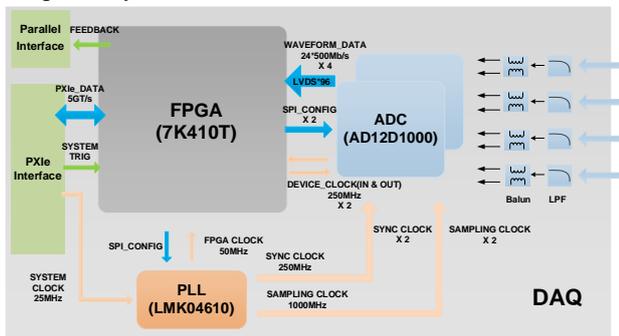

Fig. 17. Block diagram of DAQ.

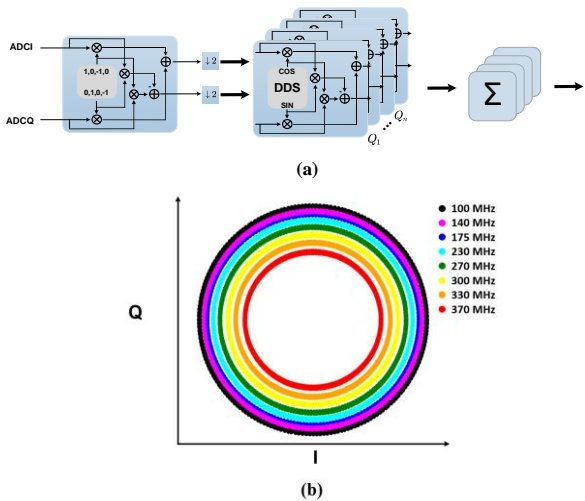

Fig. 18. (a) Structure of parallel multi-channel demodulation. (b) Test results of parallel eight-channel demodulation. Based on the testing platform as shown in Fig. 11, the AWG generates waveform consisting of eight frequency superpositions, continuously rotates the initial phase and sending it to the DAQ. DAQ completes the demodulation of 8-channel in real time, and draws the demodulated data on the IQ plane.

## 3. Bias voltage generator

The block diagram of BVG is shown in Fig. 19. The BVG is implemented with six high-precision voltage generator circuits. The circuit is implemented based on a high-precision DAC (AD5791), which provides a 20-bit resolution and 0.05 ppm/°C temperature drift. High-precision amplifier AD8676 and AD8675 is used as the reference and output buffers for AD5791. Even though precision sub-1ppm components are used, building a high-precision system is not a task that should be taken lightly. As with all precision circuits, drift of all components with temperature is a major source of error. In the prototype development of BVG, we ignored the influence of temperature drift. The test results are shown in Fig.20, and it can be seen that the output voltage is significantly affected by the temperature drift. In the subsequent design, we choose critical components such as reference (LTZ1000ach) and voltage-divider resistance with very low temperature coefficients to minimizing the drift as much as possible. Enclosing the circuit to shield the circuit from airflow would be an effective thermoelectric voltage stabilization method if further higher precision voltage output is desired.[43]

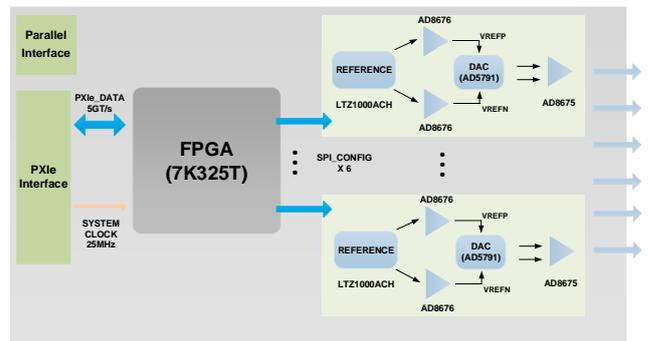

Fig. 19. Block diagram of BVG.

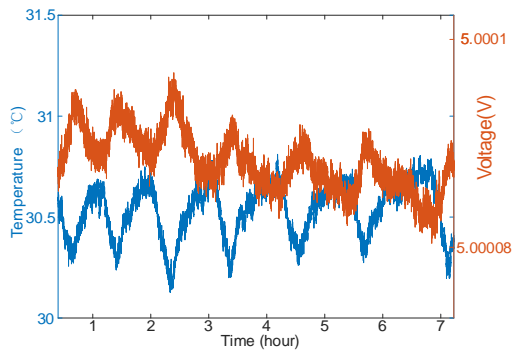

Fig. 20. Output voltage fluctuation versus temperature drift.

## DATA AVAILABILITY

The data (contain design files and HDL code) that support the findings of this study are available from the corresponding author upon reasonable request.